\documentstyle[aps,prabib,preprint]{revtex}
\begin{document}

\draft

\title{Dynamical simulation of fluidized beds --
hydrodynamically interacting granular particles}
\author{Kengo Ichiki$^{(1)}$
  \thanks{E-mail address: ichiki@cmpt01.phys.tohoku.ac.jp}
 and Hisao Hayakawa$^{(1),(2)}$
  \thanks{E-mail address: hisao@engels.physics.uiuc.edu}}
\address{$^{(1)}$Department of Physics, Tohoku University,
  Sendai 980-77, Japan\\
  $^{(2)}$Department of Physics, University of Illinois at Urbana-Champaign,
  1110 West Green Street, Urbana, IL61801-3080, USA
  \thanks{Present address}}
\date{February 28, 1995}
\maketitle

\begin{abstract}
A numerical simulation of a gas-fluidized bed
  is performed without introduction of any empirical parameters.
Realistic bubbles and slugs are observed in our simulation.
It is found that the convective motion of particles
  is important for the bubbling phase and
  there is no convection in the slugging phase.
 From the simulation results,
  non-Gaussian distributions are found
  in the particle velocities
  and the relation between the deviation from Gaussian and
  the local density of particles is suggested.
It is also shown that
  the power spectra of particle velocities obey power laws.
A brief explanation on the relationship
  between the simulation results and
  the Kolmogorov scaling argument is discussed.
\end{abstract}



\section{INTRODUCTION}
\label{sec:intro}

Recently dynamics of granular systems has attracted
 much attention among  physicists
 \cite{jaeger1992,funtai1993,bideau,thornton,funtai1994}
 as a typical object of non-equilibrium statistical physics.
For example, in vibrating beds
\cite{evesque1990,evesque1989,laroche1989,taguchi1992,gallas1992},
 collisions among particles produce convection and turbulence.
However, the fluidization of granular particles
 immersed in a fluid stream,
where the  hydrodynamic interactions are relevant,
 exhibits richer phenomena, illustrative of dynamical phase transitions
 \cite{davidson,gidaspow1986,gidaspow}.

In an experiment on fluidized beds,
 we prepare a vessel containing granular particles
 and impose a gas flow from the bottom of the vessel.
When the flow rate is small enough,
 particles do not move.
This state is known as a fixed bed.
Above a critical value of flow rate,
 the fixed bed is destabilized,
 and then  is fluidized uniformly.
At larger flow rate, the uniformly fluidized bed becomes unstable
 and bubbles appear.
Increasing the flow rate further, the bubbles become larger
 and then  become slugs
 which are horizontally spread bubbles.
For further increase of flow rate the state becomes disordered, and
  finally reaches a dilute state of particles in which is recovered
  spatial homogeneity.
These phenomena are similar to boiling of  water.
 The phase transitions in  fluidized beds, however, are not
 thermal phase transitions.
Thus the mechanism of the phase transitions in fluidized beds
 must differ from that of the boiling of  water.

These dynamical phase transitions have not been observed
 in the systems of smaller particles in flows,
 such as colloid particles
 which are much smaller than granular particle\cite{russel}.
The reason  such interesting phenomena are observed
 in granular systems may be absence of Brownian motion.
Namely, granular systems cannot reach any equilibrium states, while
colloid particles can reach. Thus, granular particles are one of interesting
subjects in nonequilibrium statistical physics.

We do not have any established model which is suitable to
 describe fluidized beds.
A modern and successful approach is simulation by
 the distinct element method (DEM)\cite{tsuji1992,tanaka1993,tsuji1993}.
The DEM is also a powerful tool
 to describe vibrating beds\cite{taguchi1992,gallas1992}.
In this approach,
 the interactions among particles are replaced by a mechanical
 model which consists of springs, dashpots and sliders.
The fluid motion is assumed to obey a phenomenological model
 where the hydrodynamic interactions
 are replaced by the coarse-grained friction
 between the particles and the fluid.
In this way, hundreds of  thousands of particles have been successfully
simulated
and a realistic motion of particles is reproduced.
However, we stress that the DEM is not independent of the experiments
because experimental
 results are used to choose parameters.

Two-fluid models are often used to describe fluidized beds
 \cite{gidaspow1986,gidaspow}.
These models allow one to understand
  macroscopic pattern formation,
  using  bifurcation analysis and
 hydrodynamic stability analysis.
In addition,
  simulations based on the two-fluid models
 reproduce realistic motions
 \cite{gidaspow,ding1990,pre-komatsu},
 and some authors indicate that
  solitons play an important role
 near the onset of the instability of uniformly fluidized beds
 \cite{sasa1992,hayakawa1993b,komatsu1993,hayakawa1994a,harris1994}.
In spite of these successful results,
  the two-fluid model contains some difficulties.
For example,
 it is difficult to choose a suitable two-fluid model\cite{batchelor1988}.
The role of particle motion is not clear
 because particles are described as a fluid.
Furthermore, the two-fluid model is supplemented
 by empirical laws for the choice of parameters.
Although two-fluid models contain a  continuous approximation,
 the simulation of two-fluid models is not easier than the DEM.

In this paper,  we perform a simulation
 based on the model which
 does not contain any empirical parameters
 except for
 the particle radius, the mass densities of the fluid and the particles,
 and the shear viscosity of the fluid.

For this purpose, we neglect the complexities of granular systems,
 which are polydispersity,
 direct interactions among the particles
 such as the Coulomb interaction
 and intermolecular forces,
 and chemical reactions induced by mixing, etc.
We only treat systems
 which contain monodisperse spheres
 with hard-core interactions among the particles
 and with hydrodynamic interactions described by the Stokes approximation.

In the next section
 we show how to construct the model and
 discuss its relevance in detail.
In Sec. \ref{sec:results},
  it is shown that
  bubbles and slugs are observed in the simulation.
In Sec. \ref{sec:analysis}
 we analyze the data obtained from the simulation.
For example, we show the distribution functions
 and the power spectra of particle velocities.
In Sec. \ref{sec:discuss},
 we briefly explain how  power laws in power spectra appear.
In Sec. \ref{sec:conclusion},
 we conclude and summarize our results.
We will present the details of our calculation method
 for hydrodynamic interactions among particles in Appendix A,
 and the treatment of fixed particles in Appendix B.

\section{SIMULATION METHOD}
\label{sec:sim-method}

In this  section, we summarize the algorithm of our simulation
 for granular particles immersed in a fluid stream.
This section consists of two parts.
The first part is devoted to the general
 perspective of the motion of particles in a fluid.
In the next part, we discuss the validity of our
 approximations.
The hydrodynamic interactions are calculated
 by the Stokesian dynamics method
 \cite{durlofsky1987a,brady1988a,brady1988b}
 which is briefly described in Appendix \ref{sec:res-matrix}.

In general,  classical particles with  mass $m$
 in a fluid obey the Langevin equation
  \begin{equation}
    m{d\over dt}{\bf U}
    =
    {\bf F}_f+{\bf F}_g
    +{\bf F}_i+{\bf F}_b,
    \label{eq:langevin}
  \end{equation}
where the velocity ${\bf U}$, the position ${\bf x}$ and
the forces ${\bf F}_l$ ($l=f,g,i,b$)
represent vectors containing $N$ particle elements as
\begin{equation}
  {\bf U}
  =
  \left[
  \begin{array}{c}
    {\bf U}^{(1)}\\
    \vdots\\
    {\bf U}^{(N)}
  \end{array}\right]
  ,\hspace{5mm}
  {\bf x}
  =
  \left[
  \begin{array}{c}
    {\bf x}^{(1)}\\
    \vdots\\
    {\bf x}^{(N)}
  \end{array}\right]
  ,\hspace{5mm}
  {\bf F}
  =
  \left[
  \begin{array}{c}
    {\bf F}^{(1)}\\
    \vdots\\
    {\bf F}^{(N)}
  \end{array}\right],
  \label{st-gp-notation}
\end{equation}
where the superscript $(i)$ represents the index of the particle.
In Eq. (\ref{eq:langevin}) there are four kinds of forces:
 ${\bf F}_f$ is the drag force from the fluid,
 ${\bf F}_g$ is the gravitational force exerted on the particles,
 ${\bf F}_i$ is the force due to direct interactions among particles,
 and ${\bf F}_b$ is the Brownian force coming from the thermal motion of
 the fluid.
Although the size distribution and the shape of particles
 are important factors in technology,
 we restrict ourselves to the motion of
 monodisperse spherical suspensions.

We define the following dimensionless quantities,
\begin{equation}
\label{eq:Pe}
  Pe = \displaystyle{6\pi\mu a^2{\cal V}\over k_BT}
\end{equation}
\begin{equation}
\label{eq:Re}
  Re = \displaystyle{\rho_fa^3{\cal V}^2\over\mu a^2{\cal V}}
  = \displaystyle{\rho_fa{\cal V}\over\mu}
\end{equation}
\begin{equation}
\label{eq:St}
  St = \displaystyle{m{\cal V}^2\over 6\pi\mu a^2{\cal V}}
  = \displaystyle{2\over 9}{\rho_p a{\cal V}\over\mu}
\end{equation}
where $Pe$,$Re$ and $St$ are respectively the P\'eclet number,
  the Reynolds number and the Stokes number.
In Eqs.(\ref{eq:Pe})-(\ref{eq:St}), $a$, $\rho_p$, $\rho_f$ and
$\mu$ are the particle radius, the mass densities of
 the particle and the fluid, and the shear viscosity of the fluid respectively,
$k_B$ is the Boltzmann's constant.
${\cal V}$ is the characteristic velocity, where we choose
the sedimentation rate of one particle
\begin{equation}
  U_0=\frac{m \tilde g}{6\pi \mu a}=\frac{2 a^2 \rho_p \tilde g}{9 \mu}
\end{equation}
 with the effective gravitational acceleration
 $\tilde g=g(\rho_p-\rho_f)/\rho_p$.
These dimensionless numbers actually have definite meanings.
The P\'eclet number is the ratio of work done by the drag force
over the size of a particle
$(6\pi\mu a^2{\cal V})$ to thermal energy $(k_BT)$.
The Reynolds number is well known as the ratio
of inertia to  drag.
The Stokes number represents the relative importance of
 the kinetic energy of particles $(m{\cal V}^2)$ to work done by
the  drag force.
It is also recognized as the ratio of the time scales,
 \begin{equation}
   St = {T_{r}\over T_{p}},
   \label{eq:st-time}
 \end{equation}
 where $T_{r}=m/6\pi\mu a$ is
 the relaxation time of particle velocity due to the drag force,
 and $T_{p}=a/{\cal V}$ is the passing time of the particle scale $a$
 with the velocity ${\cal V}$.

The Froude number, which has been widely used on this problem,
 is given by
\begin{equation}
  Fr
  =
  {{\cal V}^2\over{\tilde g}L},
\end{equation}
 with $L$ being the linear size of system.
The meaning of the Froude number is
 the ratio of the kinetic energy to the gravitational potential, and
$Fr$ is proportional to $St$ if we adopt ${\cal V}=U_0$.
For later convenience, we also introduce the effective Reynolds number of
particles as
 \begin{equation}
  Re_{(p)}
  \equiv
  {{\cal V}L\over \nu_p}
  = {9L\over 2a}St,
   \label{eq:fr-st}
 \end{equation}
where $\nu_p=\mu/\rho_p$.
The Stokes number is related to the particle size,
 while the Froude number and the effective Reynolds number $Re_{(p)}$
are related to the system size.
The role of $Re_{(p)}$ will be discussed in Sec. \ref{sec:discuss}.
Thus $Pe$,$Re$ and $St$ are fundamental and independent parameters to
 determine the motion of particles.

Let us consider the explicit form of forces in (\ref{eq:langevin})
 and discuss their relevance.
The gravitational force ${\bf F}_g$ acting in the $z$ direction
 is simply given by
\begin{equation}
  {\bf F}_g=-m\tilde{g}{\bf E}_z,
\end{equation}
where
 ${\bf E}_z$ is the generalization of unit vector ${\bf e}_z$
 as in (\ref{st-gp-notation}).
The direct interaction among particles
 ${\bf F}_i$ is assumed to be due to hard-core interactions.
Therefore it can be treated as exchanging velocities
 during elastic collisions among particles.
We neglect the random force ${\bf F}_b$,
  because in this paper we treat the system where
  the P\'eclet number is large enough.
The validity of $Pe\gg 1$ will be discussed later.

The most relevant and complicated force is ${\bf F}_f$,
 which is determined by the Navier-Stokes equation
\begin{equation}
  \rho_f\left\{\partial_t{\bf u}
  +({\bf u}\cdot\nabla){\bf u}\right\}
  =
  \mu\nabla^2{\bf u}-\nabla p,
\end{equation}
with the incompressible condition
\begin{equation}
  \nabla\cdot{\bf u}
  =
  0.
\end{equation}
The incompressibility is valid even if
 the fluid is air, when the particle motion is much slower than
 the sound velocity of fluid.
When $Re\ll 1$, the Navier-Stokes equation is reduced to
 the Stokes equation\index{Stokes equation}
\begin{equation}
  -\mu\nabla^2{\bf u}+\nabla p=\vec{0}.
\end{equation}
In this case the particle velocity ${\bf U}$ is
connected with the force induced by the particles on the fluid
${\bf F}_p$ in the following linear relation \cite{happel},
\begin{equation}
  {\bf F}_p
  =
  {\sf R}\left({{\bf x}}\right)\cdot
  \left({\bf U}-{\bf u}^\infty\right),
  \label{eq:resistance}
\end{equation}
 where ${\bf u}^\infty$ is the fluid velocity in the absence of particles
 and ${\sf R}$ is the $3N\times 3N$ resistance matrix
 which depends only on the particle configuration ${\bf x}$.
The details of the construction of ${\sf R}$ are described
 in Appendix \ref{sec:res-matrix}.
The force ${\bf F}_p$ induced by the particle is related to
 ${\bf F}_f$ by
\begin{equation}
  {\bf F}_p=-{\bf F}_f.
\end{equation}
In later discussions,
 we assume $Re\ll 1$.
As we will show,
 this approximation is valid for the motion of relatively small particles.

In the case of $Re\ll 1$ and $Pe\gg 1$
 and with hard-core interactions,
 the Langevin equation (\ref{eq:langevin}) can be reduced to
  \begin{equation}
    St
    {d\over d\hat{t}}\hat{{\bf U}}(\hat{t})
    =
    -\hat{\sf R}({\bf x}(\hat{t}))\cdot
    \left(\hat{{\bf U}}(\hat{t})-\hat{{\bf u}}^\infty\right)
    -{\bf E}_z.
    \label{eq:langevin-nd}
  \end{equation}
Here we scale the velocities by ${\cal V}$,
  the length by the radius $a$ and
  the resistance matrix by the drag factor $6\pi\mu a$.
 We denote the dimensionless value of ${\bf U}$ as $\hat{\ {\bf U}}$.
 From (\ref{eq:langevin-nd})
  it is obvious that
  the Stokes number $St$ is an important parameter for our model.
The role of $St$ is also pointed out for
  dilute monodisperse suspensions with low Reynolds number
  \cite{koch1990,koch1992b}.
For vibrating beds in which the particle inertia is dominant and
  the hydrodynamic interaction is negligible,
  $St$ should be large while
  for liquid-fluidized beds
  $St$ is small.

In our simulation, we need to integrate Eq.(\ref{eq:langevin-nd})
  for a small time interval numerically.
We divide $\hat{\bf U}(\hat{t})$ into two parts,
\begin{equation}
  \hat{{\bf U}}(\hat{t})=\hat{{\bf U}}_0+\hat{{\bf U}}_1(\hat{t}).
  \label{eq:v01}
\end{equation}
$\hat{{\bf U}}_0$ is determined by
\begin{equation}
  \vec{0}
  =
  -\hat{\sf R}({\bf x})\cdot
  \left(\hat{{\bf U}}_0-\hat{{\bf u}}^\infty\right)
  -{\bf E}_z.
  \label{eq:v-term}
\end{equation}
It is obvious that
  ${\bf U}_0$ is the same solution as that for $St\ll 1$.
Although $\hat{{\bf U}}_1(\hat{t})$ is determined by
\begin{equation}
  St
  {d\over d\hat{t}}\hat{{\bf U}}_1(\hat{t})
  =
  -\hat{\sf R}({\bf x})\cdot
  \hat{{\bf U}}_1(\hat{t}),
  \label{eq:vdif-1}
\end{equation}
  we adopt the simplest form
  for computational efficiency,
\begin{equation}
\label{eq:crucial}
  \hat{{\bf U}}_1(\hat{t})
  =
  \hat{{\bf U}}_1(0)\ \exp\left(-{\hat{t}\over St}\right),
\end{equation}
  which is the solution of (\ref{eq:vdif-1}) on the assumption of
 $\hat{\sf R}\simeq {\sf I}$,
 where ${\sf I}$ is the unit tensor.
 From the initial condition
  $\hat{{\bf U}}(0)=\hat{{\bf U}}_0+\hat{{\bf U}}_1(0)$,
we thus obtain
  \begin{equation}
    \hat{{\bf U}}(\hat{t})=\hat{{\bf U}}_0+
    (\hat{{\bf U}}(0)-\hat{{\bf U}}_0)\exp\left(-{\hat{t}\over St}\right).
    \label{eq:time-evol-vel}
  \end{equation}
This solution shows us that
  the particle velocities are damped
  from the initial value $\hat{{\bf U}}(0)$
  to the terminal velocity $\hat{{\bf U}}_0$.
The simplification in Eq.(\ref{eq:crucial}) is crucial.
We expect, however, the error
 from this simplification is small
 when we choose a small enough time interval
 for the numerical integration,
 because the resistance matrix is calculated as a
 function of particle configuration at each step.

Let us estimate the values of the dimensionless parameters.
We adopt $\rho_p =2.5 {\rm [g\ cm^{-3}]}$
 which is a typical value of glass beads.
Fluids are assumed to be air
 ($\rho_f = 1.2\times 10^{-3} {\rm [g\ cm^{-3}]}$ and
 $\mu = 1.82\times 10^{-4} {\rm [g/cm\ sec]}$)
 and water ($\rho_f = 1.0 {\rm [g\ cm^{-3}]}$ and
 $\mu = 1.0\times 10^{-2} {\rm [g/cm\ sec]}$) at room temperature.
Substituting $k_B=1.38\times 10^{-16}{\rm [erg\ K^{-1}]}$,
  $T=293{\rm [K]}=20[ ^\circ C]$ and
  $g=981 {\rm [cm\ sec^{-2}]}$,
  we obtain Table \ref{tab:param} and Fig. \ref{fig:param}.
We also show the characteristic times,
  $T_{r}$ and $T_{p}$, in (\ref{eq:st-time}).

The lines in Fig. \ref{fig:param} connect the points
  where dimensionless parameters
  are equal to unity for air and water.
Therefore, there is no meaning in the vertical axis.
In the left of the line of $Pe=1$ in Fig. \ref{fig:param},
  the random force is dominant
  and the particles can be regarded as typical colloidal suspensions.
On the other hand, the particles to the right of
 $Pe=1$ can be regarded as typical granular particles
 where the hydrodynamic interaction among particles dominates
 the Brownian force.
In the left region of $Re=1$,
 the fluid motion can be described by the Stokes equation.
On the other hand, in the right  of $Re=1$,
 the inertia term in the Navier-Stokes equation is important.
In the left region of $St=1$ where $T_{r}<T_{p}$,
 we can neglect the inertia effect of particles.
Thus, the particles move with their terminal velocities.
In the right of $St=1$ where $T_{r}>T_{p}$,
 the effect of particle inertia exceeds the drag from the fluid.
Therefore, collisions among the particles are important.

We choose the particle radius as $10 \mu{\rm m}$
 because we are interested in the case of $Pe\gg 1$ and $Re\ll 1$.
In this case the dimensionless parameters have the values
 presented in Table \ref{tab:param}
 for air and water.
As anticipated in Fig. \ref{fig:param},
 the difference between  air and  water appears in the value of
 the Stokes number.
We now interested in the phenomena driven by the drag force,
 which is the behavior in the time scale $T_{p}$.
Therefore,
for gas-fluidized beds,
 the collisions among particles are not negligible,
 while for liquid-fluidized beds,
 the collisions are not important.
We focus on gas-fluidized beds in this paper.
We will discuss the properties of liquid-fluidized beds
 elsewhere.

We notice that the particle radius $a\simeq 10^{-3}$[cm]
  belongs to the group C in the classification by Geldart
  \cite{geldart1973,gidaspow}
  in which all cohesive powders are difficult to fluidize.
This difficulty in experiments arises
  because direct interparticle forces dominate
  the hydrodynamic drag force.
If the interparticle forces, except for the hard-core interactions
  can be removed,
  a collection of small particles must be fluidized as will be shown.

Finally we summarize the adopted assumptions.
   \begin{enumerate}
  \item the diameter of all particles is identical.\label{item:monodisp}
  \item the interaction among particles is described by hard-core interactions.
  \item the thermal motion is negligible. \label{item:no-random}
  \item the inertia of the fluid is negligible. \label{item:stokes-approx}
  \end{enumerate}
Assumptions \ref{item:no-random} and \ref{item:stokes-approx} are valid
  when the particle radius is between $1[\mu{\rm m}]$ and $10[\mu{\rm m}]$
  in both air and water.

\section{SIMULATION RESULTS}
\label{sec:results}

In this section, we collect three typical results of simulations.
In many of our simulations,
 the configuration of particles is restricted to lie in a plane
 parallel to the direction of gravity
 for efficiency of calculation,
although  periodic boundary conditions are also applied in the direction
 perpendicular to this plane.
Thus, in such cases,
 particles are influenced by
 three-dimensional hydrodynamic interactions.

At first
 we show the result of our simulation
 using rectangular cells with periodic boundary conditions
 in which the ratio of the height to the width is large compared
 with unity.
We observe stable slugs which move upward in this simulation
 (Fig. \ref{fig:slug}).
This result is also observed in the three-dimensional (not monolayer)
 simulation with slender cells as well (Fig. \ref{fig:slug-3d}).
 From these results,
 particles in the dilute region relatively fall down,
 so that in the denser region the sedimentation velocity is smaller.
This tendency agrees with the standard theory
 \cite{batchelor1972a,brady1988c,prepri-HI94}
 and the experiments of sedimentation\cite{russel,davis1985}.
In our simulation for slugs
 the configuration of particles
 is nearly closed packed and particles are immobile in concentrated regions.
On the other hand in the dilute region,
 particles monotonically fall down.
We cannot observe any convection of particles
 in this simulation.
Thus slugs can be produced by a pseudo one-dimensional motion
 of particles, after the uniform state has become unstable.

Next
 we show the result of our simulation
 using square unit cells with  periodic boundary conditions.
 Figure \ref{fig:square} is a typical snapshot.
In  contrast to the previous result,
 we see that
  particles in the dilute region float up,
  while particles in the dense region fall down.
Thus particle convection exists around the dilute region,
 which may be regarded as a bubble.
This kind of convection inside bubbles has been observed in experiments
  \cite{private-mori}.
Thus, we infer that convection is important to create bubbles.
This bubble, however, is not stable and
  it will disappear soon afterwards.
We can also see periodic birth and death processes of bubbles.

We have also performed simulations
 introducing fixed particles in our system.
The treatment of fixed particles is described
  in Appendix \ref{sec:meth-fixed}.
In real systems,
  particles are settled in a vessel.
For this purpose,
 we introduce particles fixed at the bottom of the unit cell.
In our simulation, the fixed particles are placed horizontally,
 spaced as far as 3.5 radius apart.
We perform this simulation as follows.
At first we randomly position free particles and
 allow them to drift to the bottom under the influence of gravity.
Thus the free particles fall down and produce a fixed bed.
Next we inject the flow upward with the velocity ${\bf u}^\infty$.
The result of our simulation after the injection of flow
 is shown in Fig. \ref{fig:fixed}.
We notice that the unit cell drawn in Fig. \ref{fig:fixed}
 is connected with its mirrors in all directions
 as in the previous figures.

 From Fig. \ref{fig:fixed}, we observe that at first
 the fixed bed floats like a single cluster,
 but the cluster becomes unstable.
Then the fixed bed becomes a fluidized bed.
In this fluidized bed, we observe the formation of bubbles
 periodically at the bottom, which float up,
 and travel through the bed.
We also observe  particle convection around  bubbles.
This kind of bubble formation from fixed beds
 is similar to that in the real experiments\cite{private-mori}
 and large size simulation
 \cite{davidson,tsuji1992,tanaka1993,tsuji1993}.

Figure \ref{fig:fixed-variance} shows the standard deviation of
 the particle velocities and
Figure \ref{fig:fixed-density} shows the number of particles
 in the region at $8/25$ of the cell height above the bottom.
These figures show that
 bubble formation occurs periodically at the peaks of the standard deviation.

At the end of this section, we summarize important characteristics in
fluidized beds.
When we compare the results  in Figs. \ref{fig:slug} and \ref{fig:square} or
 \ref{fig:fixed},
 we observe that the particle motion has different characteristics.
In Fig. \ref{fig:square},
 particles float up in dilute regions and
 fall down in concentrated regions.
This tendency can be understood as follows. The concentrated regions can be
regarded as clusters, where flow cannot penetrate inside clusters.
Thus clusters may have fast sedimentation velocity as in the definition of
${\cal V}\propto a^2$ with the radius $a$. On the other hand,
slugging motion is understood from  the standard theory of sedimentation of
homogeneous suspensions, where dilute regions have larger sedimentation rate
than
dense regions.
In Fig. \ref{fig:slug} we see no isolated clusters, and particle distribution
for horizontal direction is almost uniform.
If particles are dispersed uniformly, fluid flow feels larger drags
from  many particles than that from small particles.
In other word,
  the difference between the bubbling phase and the slugging phase
  is whether or not the convection of particles exists.
To produce convections, we need both characteritics of
clusterings and sedimentation.
Thus, the particle motions in real systems
 are determined by complex combinations of
two different characteristics.

\section{ANALYSIS}
\label{sec:analysis}

In this section, we analyze
  the data obtained from our simulation.
At first we discuss the velocity distribution functions (VDFs),
  where not only  Gaussian-like but non-Gaussian distributions
  are observed
  and close relationship between the deviation from Gaussian
  and the local density of particles is indecated.
Next
  we discuss the power spectra obtained from the data,
  which suggest the existence of tails obeying power laws.
These power law tails  may correspond to those
 reported by Taguchi\cite{taguchi1993c},
 and taken
 as an evidence for powder turbulence \cite{taguchi1995}.

We display VDFs in our simulation
 for systems with the square periodic cells
 in Fig. \ref{fig:veldist-square}.
 From this figure,
  it is obvious that the VDFs are near Gaussian but anisotropic.
In this case, the  vertical VDF has two branches, each
  obeying a different Gaussian distribution.
This  transmission in the VDF has also been observed
  in the simulation of the two-fluid model\cite{pre-komatsu}.
The anistropic poperty of VDF
 indicates that the system does not have any local equilibrium.

For the systems with fixed particles,
 the form of VDFs is similar to an exponential one
 (Fig. \ref{fig:veldist-fixed}).
It is interesting that
  the introduction of fixed particles produces an exponential VDF.
In this system
 as shown in Fig. \ref{fig:fixed-variance},
 active states which has large standard deviation of particle velocities
 and inactive states of motion of particles emerge in turn.
 From Figs. \ref{fig:fixed-variance} and \ref{fig:fixed-density},
 we indicate that the active state corresponds to the appearance of a bubble
 and in the inactive state contains is no bubble.
To see the quantitative difference between two states,
 we investigate VDF in each state separately.
Here we define the active state as the time region
 where the standard deviation of particle velocities is more than $0.4U_0$,
 and the inactive state as the period where the standard deviation is
 less than $0.3U_0$.
Figure  \ref{fig:fixed-vdf-hard-soft} is
 the  VDFs for the active and the inactive state.
 From this figure, VDF in the inactive state is close to an
 exponential distribution,
 while that in the active state is a Gaussian distribution.

For the systems with slugs,
 VDFs are far from Gaussian futher and
 also different from simple exponential distribution
 (Fig. \ref{fig:veldist-slug}).
We attempt to fit them to t-distribution,
 \begin{equation}
   f(U)
   \sim
   \left(1+aU^2\right)^{-b},
   \label{eq:t-distrib}
 \end{equation}
 where $a$ and $b$ are parameters.
This t-distribution is observed by
  Taguchi and Takayasu\cite{taguchi1994b} for VDFs in vibrating beds
  and by Sinai and Yakhot\cite{sinai1989} for passive scalars in turbulence.
 From this figures, although the horizontail VDF can be fitted by
  the t-distribution,
  the vertical VDF is far from any t-distribution.
However the tail for negative $U_y$
 can be understood by the following simple picture.
In the slugging state,
  most of particles are included in the dense region
  and only a few particles are falling in the slug.
Here we assume that
  falling particles start from zero relative velocity to the dense region.
 The particles are accelerated by the gravity then
 they collide with the top of the dense region
  and join in it.
If this assumptions are valid,
  the VDF of falling particles is uniform
  in the range between the initial and the terminal falling velocities,
  which may correspond to the  tail for $U_y$
observed in Fig. \ref{fig:veldist-slug} .
To investigate the VDF in the dense region,
  we substract the uniform distribution from the original VDF
  (Fig. \ref{fig:veldist-slug} (b))
  in the range of $-0.4<\hat{U}<-0.115$.
Here the probability of the uniform distribution is approximated
  by the original VDF in the range of $-0.7<\hat{U}<-0.4$,
  and $\hat{U}=-0.115$ is the location of the peak of the original VDF.
The resultant VDF is shown in Fig. \ref{fig:vdf-slug-sub}
  in the range of $-0.4<\hat{U}<0.3$.
We can fit this well by (\ref{eq:t-distrib}) for each side of the peak,
although
 there is still anisotopy in the figure. Thus, $t-$ distribution
seems to be applicable to our system.

 From our results of VDFs,
  we may indicate that
  non-Gaussian properties are closely related to
  the local density of particles.
Because in relatively dilute case such as the system with square unit cells
  and the active state in the system with fixed particles,
  VDFs are close to a Gaussian,
  while in the dense case such as the inactive state in the system
  with fixed particles and the system which has slugs,
  VDFs are far from Gaussian.
These two characteristics may be understood as follows.
In active states,
 the density of particles are relatively low.
 As a result, particles can move almost freely without influence of
lubrication effects.
 Then collisions among particles occur at random,
 and inelastic effects from viscous terms which is mainly from the lubrication
force may be suppressed.
On the other hand, in inactive states,
 the density of particles is extremely high,
 and the lubrication effect also becomes important.
Therefore, the inelastic effects dominate the randomness to produce
 Gaussian distributions.

It is interesting that we observe non-Gaussian, exponential-like,
 VDFs in our systems.
Non-Gaussian probability distributions are observed in various systems.
In fluid turbulence
 non-Gaussian distributions are found in
 the probability distribution functions
 of velocity differences\cite{vanatta1970}
 and passive scalars\cite{sano1989,castaing1989}.
In simulations of vibrating beds,
 VDFs of particles are found to be
 described by a non-Gaussian VDFs\cite{taguchi1994b}.
For the simple models consisting of hard-spheres,
 VDFs also show non-Gaussian distribution\cite{pre-taguchi1995,jarzynski1993}.
In astronomy non-Gaussian VDFs are also reported\cite{pre-miesch1994}.

We also indicate that
  in the region obeying non-Gaussian VDFs,
  the kinetic temperature which  is defined through
  the deviation of the Gaussian VDF cannot be used.
Although it is possible to introduce the granular temperature
from the different context, we need to be aware of diffrence between
this granular temperature and the usual kinetic temperature.

Next, we investigate the power spectra in frequency $E(\omega)$
  defined through
\begin{equation}
  E(\omega)
  =
  {1\over N}
  \sum_\alpha^N
  <\tilde{{\bf U}}^\alpha(\omega)
  \cdot{\tilde{\bf U}}^{\alpha *}(\omega)>,
  \label{eq:eofw}
\end{equation}
where
\begin{equation}
  \tilde{\bf U}(\omega)
  =
  \int_{-\infty}^\infty dt\ e^{-i\omega t}{\bf U}(t).
\end{equation}
The results shown in Figs. \ref{fig:pow-t-square},
 \ref{fig:pow-t-slug} and \ref{fig:pow-t-fixed}
 are obtained by a standard fast Fourier transform routine
 with the Parzen window\cite{numerical-recipes}.
All of these figures indicate that there are three regions;
  the spectra seems to be white in  low $\omega$,
 there are some peaks in ther middle,
  and the spectrum obeys a power law in high $\omega$.
To understand the mechanism to make three regions,
we consider the following three characteristic frequencies,
\begin{equation}
  \omega_{L}\equiv 2\pi{\bar{U}\over L},
\end{equation}
\begin{equation}
  \omega_{p}\equiv {2\pi\over T_{p}},
\end{equation}
\begin{equation}
  \omega_{r}\equiv {2\pi\over T_{r}},
\end{equation}
 where we adopt  the average particle velocity relative to the fluid
 for the systems without fixed particles
 and the induced flow rate for the system with fixed particles as $\bar{U}$.
$T_{p}$ and $T_{r}$ introduced in (\ref{eq:st-time}) are
 the passing time of the particle scale and
 the relaxation time respectively.
These frequencies are also shown in the figures.
 From these figures,
$\omega_L$ and $\omega_r$ seem to correspond to boundaries of three regions;
  the white spectrum region, the region with some peaks,
  and the region obeying power law.
The peak near $\omega/\omega_0\simeq 13$ in Fig. \ref{fig:pow-t-fixed}
 corresponds to the frequency of bubble formations,
 where $\omega_0=2\pi/2048\times 10^{-4}$[sec]is
 the smallest frequency produced from our entire simulation.
In the higher frequency range,
  all of these figures show the existence of a power law
  $E(\omega)\sim \omega^{\beta}$ between $\omega_r$ and $\omega_p$
  whose exponent is fluctuated between $-1.49$ and $-1.63$.
This is not far from the Kolmogorov spectrum $E(\omega)\sim \omega^{-5/3}$
  in fluid turbulence.
About the power laws, we will discuss futher in the next section.

\section{DISCUSSION}
\label{sec:discuss}

Now we discuss the origin of the power laws observed in the power spectra,
  which are similar to Kolmogorov scaling \cite{kolmogorov1941}.
Taguchi\cite{taguchi1993c} also observed the Kolmogorov-like scaling
 in vibrating beds.
Therefore we need to clarify whether the power law observed in our simulations
 is the same as that of Taguchi.

Our system contains an energy source on the scale of the lowest wave number,
 and the energy is dissipated in the fluid on the scale of the highest wave
number.
Vibrating beds have also common feature.
Therefore,
 we might expect to have a cascade process, as  assumed by Kolmogorov and
proposed by Richardson,
 where the energy dissipation rate $\epsilon$
 determines the statistical properties of small particle motion.
For this purpose,
 we assume that the motion of particles is determined by only
 the energy dissipation rate $\epsilon$ and
 the viscousity $\mu$.
As discussed in Sec. \ref{sec:sim-method},
  the particle motion is described
  by only one relevant dimensionless parameter,
  the Stokes number, $St = 2\rho_p a{\cal V}/9\mu$
  in the case of $Pe \gg 1$ and $Re \ll 1$.
Although the Stokes number may be regarded as the effective Reynolds number
  for the particle fluid,
  it contains only the energy dissipation length scale.
Instead of $St$
 it is convenient to use
the effective Reynolds number for particle fluid
$Re_{(p)}\equiv {\cal V}L/\nu_p=9L St/2a$ defined in Eq.(8).
 From dimensional analysis,
 we obtain the following scaling for the energy spectrum
 \begin{equation}
   E(k)
   =
   \epsilon^{1/4}\nu_p^{5/4}   F\left({k\over k_d},Re_{(p)}\right),
 \end{equation}
 where $F(x,y)$ is a dimensionless function,
 $k_d$ is the dissipation scale which is the maximum of
  $a^{-1}$ and the Kolmogorov scale $\epsilon^{1/4}\nu_p^{-3/4}$.
In the region of high wave numbers
 we might guess that there is a balance between energy injection and
cascade process, where the dissipation  is not important.
Therefore, we obtain
\begin{equation}
  E(k)
  =
  \epsilon^{2/3}k^{-5/3}.
  \label{eq:Kol-scale-2}
\end{equation}
This result is equivalent to that by Kolmogorov\cite{kolmogorov1941}.

In order to ensure the scaling ansatz,
 we need to impose the condition
 \begin{equation}
   Lk_d
   \sim
  Re_{(p)}^{3/4}
   \gg 1.
   \label{eq:scaling-ansatz}
 \end{equation}
Equation (\ref{eq:scaling-ansatz}) means that
 the inertia spectrum can be observed in our system (Table \ref{tab:froude}).
Although there is an ambiguity in choosing  $L$,
  it is interesting that the values of $Re_{(p)}^{3/4}$ correspond to
  the strength of the non-Gaussain VDFs.
We also expect the cutoff $a^{-1}$ is larger than
 the dissipation scale $\epsilon^{1/4}\nu_p^{-3/4}$,
 because we do not observed any dissipation range
 (see also \cite{taguchi1993c}).
We also comment on the reason to observe turbulent
 charactersitics in relatively small $Re_{(p)}$,
 where in usual fluid, the value of $Re_{(p)}$ in Table \ref{tab:froude}
 is not enough large to show turbulent charcateristics.
We guess that inelatic scatterings of particles
 in our system enhance chaotic characteristics,
 while the molecules in pure fluid only have elastic scatterings.
We can use this argument to vibrating beds,
 where $\mu$ is replaced by the effective viscosity of particle flow.

We only discuss the spectrum for $E(k)$
  and not for our observed $E(\omega)$.
Although the general relationship between $E(k)$ and $E(\omega)$ is unclear,
  Viecelli \cite{viecelli1991} demonstrates
\begin{equation}
  E(\omega)
  \sim
  \omega^{-5/3}.
\end{equation}
for the inertia range
in both Euclerian and Lagrangian coordinate frames can be derived
from (\ref{eq:Kol-scale-2}) with the assuption of space-time symmetry.
His argument is independent of the basic equations and assumes
that the system can be charcaterized only through
the kinetic viscoisty and $\epsilon$ as in our argument.
 Therefore, we may expect that his argument
is applicable to our case.

We also comment on the reason that
 we observe the Kolmogorov like scaling even in two-dimensional systems.
In our case the enstrophy is not a conserved quantity in two dimensional case.
Therefore it is not surprizing that we observe the Kolmogorov-like scaling
 in two-dimensional systems,
 because the derivation of (\ref{eq:Kol-scale-2})
 does not contain any information about
 the spatial dimensionality.

\section{CONCLUSION}
\label{sec:conclusion}

In this paper, we have succeeded in  simulating  granular systems with
 fluid flow without any empirical parameters.
Our simulation reproduces realistic slugs and bubbles.
According to the results of our simulation,
 we confirm that  particle convection is important for bubble
 formation.
 From our simulation we observe several velocity distribution functions,
  Gaussian and non-Gaussian or exponential distribution.
We also observe  power spectra obeying power laws.
We briefly explain the relationship between our observed spectra and
  Kolmogorov scaling.
In this sense,
 our paper has confirmed the universality of the powder turbulence
 proposed by Taguchi\cite{taguchi1995}.

Before closing this paper,
 we comment on our approximation in using  the Stokes equation for fluid flow.
In real experiments and simulations by engineers,
 the particle radius, typically $a\simeq 10^{-1}$ [cm],
 is often much larger than that we assume ($a\simeq 10^{-3}$ [cm]).
This is because the direct interaction among particles except for
  hard-core interactions dominates all other forces,
  and it is difficult to observe fluidized beds in such small particles
  \cite{gidaspow,geldart1973}.
For the fluidized beds with such large particles,
 we need to consider the advection term in the  fluid motion
 and the drag which should contain a term in proportion to
 the square of the difference of velocities
 between fluid and particles \cite{gidaspow}.
To simulate this system without introduction of empirical laws
 is almost impossible.
We, however, have shown that
 these kinds of complexities are irrelevant for the formation of bubbles
 and slugs.
The essence of the physics in fluidization can be understood
 when we drop irrelevant complexities
 such as  direct interparticle forces
 and the advection term in fluid flows.

\section*{ACKNOWLEDGEMENTS}

The authors thank fruitful discussions with
  T.S. Komatsu, Y-h. Taguchi, S. Sasa, H. Takayasu and H. Nishimori.
We also appreciate the useful comments by H.J. Herrmann and A.S. Sangani, and
are grateful to N. Goldenfeld for his critical reading and comments.
One of the authors (K.I.) thanks J.F. Brady for explaining the
 details of his calculations to the author.
This work was partially supported by the Foundation of
 Promotion for Industrial Science and the U.S. National Science Foundation
through grant number NSF-DMR-93-14938.

\appendix

\section{Hydrodynamic interaction}
\label{sec:res-matrix}

In this section, we explain how to treat
hydrodynamic interactions.
When we adopt the Stokes approximation,
 the calculation of the hydrodynamic interactions
 is equivalent to constructing the resistance matrix.
Our aim is to calculate the resistance matrix with
 acceptable accuracy taking into account computational efficiency.
We thus adopt the Stokesian dynamics developed by
 Brady and his coworkers  \cite{durlofsky1987a,brady1988a,brady1988b}
 to describe colloidal dispersions.
They distinguish hydrodynamic long-range interactions
 from the lubrication force which represents short range hydrodynamic
 repulsive interactions.
For the long-range part, we use the multipole expansions,
 while for the lubrication part, we use the exact solution of
 two-body problem \cite{jeffrey1984,kim1985} as
\begin{equation}
  \left[
  \begin{array}{c}
    {\bf F} ^{(1)} \\
    {\bf F} ^{(2)} \\
      \end{array}\right]
  = {\sf  R }_{2B}
  \cdot
  \left[
  \begin{array}{c}
    {\bf U}^{(1)} - {\bf u}^{\infty}\\
    {\bf U}^{(2)} - {\bf u}^{\infty}\\
  \end{array}\right],
\end{equation}
where ${\bf u}^{\infty}$ is the fluid velocity without particles.
The lubrication part is separated from the exact solution as
\begin{equation}
  {\sf  R }^{lub}_{2B} = {\sf  R }_{2B}
  - \left( {\sf  M }^\infty_{2B}\right)^{-1} ,
\end{equation}
where $ {\sf  M }^\infty_{2B}$ is the Rotne-Prager tensor\cite{rotne1969},
 which represents the long-range interaction.
Let ${\sf R}^{lub}$ be the linear combination of all possible pairs of
${\sf  R }^{lub}_{2B}$.
Then, an approximate resistance matrix is represented by
\begin{equation}
  {\sf R}
  =
  \left( {\sf M}^\infty \right)^{-1} + {\sf R}^{lub}.
\end{equation}
The validity of this method has been confirmed in various
numerical experiments for colloid systems.
This idea is also valid in the theoretical calculation of
 the sedimentation rate \cite{prepri-HI94}.

For the long-range interaction,
  we apply periodic boundary conditions to describe the system
  containing infinite number of particles.
For simplicity, we consider only the contribution of force.
Note that higher order corrections from torque
 have been discussed by Brady {\it et al.}\cite{brady1988b}.
Beenakker \cite{beenakker1986} has shown that the mobility matrix
 can be written   Ewald's summation  as
  \begin{eqnarray}
    6\pi\mu a
    (U^\alpha_i - u^\infty_i)
    &=&
    F^\alpha_i
    +\sum_\gamma\mathop{{\sum}'}_{\beta =1}^NM^{(1)}_{ij}
    ({\bf x}^\alpha-{\bf x}^\beta +{\bf r}_\gamma )F^\beta_j
    \nonumber\\
    &&
    +{1\over V}\sum_{\lambda\neq 0}\sum_{\beta=1}^N
    \cos\left({\bf k}_\lambda\cdot ({\bf x}^\alpha -{\bf x}^\beta )\right)
    M^{(2)}_{ij}({\bf k}_\lambda)F^\beta_j
    -M^{(2)}_{ij}({\bf r}=\vec{0})F^\alpha_j,
    \label{eq:sim-ewald}
  \end{eqnarray}
where $N$ is the number of particles in the unit cell
 and $M^{(1)}_{ij}({\bf r})$, $M^{(2)}_{ij}({\bf k})$
 and $M^{(2)}_{ij}({\bf r}=\vec{0})$ are respectively given by
\begin{eqnarray}
    M^{(1)}_{ij}({\bf r})
    &=&
    {\rm erfc}(\xi r)\left\{
    \delta_{ij}\left(
    {3\over 4}{a\over r}+{1\over 2}{a^3\over r^3}\right)
    +\hat{r}_i\hat{r}_j\left(
    {3\over 4}{a\over r}
    -{3\over 2}{a^3\over r^3}\right)\right\}
    \nonumber\\
    &&
    +{e^{-\xi^2r^2}\over \sqrt{\pi}}\left[
    \delta_{ij}\left(
    4a^3\xi^7r^4
    +3a\xi^3r^2
    -20a^3\xi^5r^2
    -{9\over 2}a\xi
    +14a^3\xi^3
    +a^3\xi{1\over r^2}\right)\right.
    \nonumber\\
    &&\left.
    +\hat{r}_i\hat{r}_j\left(
    -4a^3\xi^7r^4
    -3a\xi^3r^2
    +16a^3\xi^5r^2
    {3\over 2}a\xi
    -2a^3\xi^3
    -3a^3\xi{1\over r^2}\right)\right],
    \label{eq:sim-ewald-1}
\end{eqnarray}
\begin{equation}
    M^{(2)}_{ij}({\bf k})
    =
    6\pi a
    \left(\delta_{ij}-\hat{k}_i\hat{k}_j\right)
    {1\over k^2}\left(1-{1\over 3}a^2k^2\right)
    \left(1+{k^2\over 4\xi^2}+{k^4\over 8\xi^4}\right)
    e^{-{k^2\over 4\xi^2}},
    \label{eq:sim-ewald-2}
\end{equation}
and
\begin{equation}
    M^{(2)}_{ij}({\bf r}=\vec{0})
    =
    {\delta_{ij}\over\sqrt{\pi}}
    \left(
    6a\xi-{40\over 3}a^3\xi^3\right).
    \label{eq:sim-ewald-20}
\end{equation}
Here ${\rm erfc}(x)$ is the complimentary error function given by
\begin{equation}
  {\rm erfc}(x)
  =
  {2\over\sqrt{\pi}}
  \int_x^\infty {\rm exp}(-z^2)dz,
\end{equation}
and $\xi$ with units of inverse length
 is an arbitrary parameter
 which we choose to minimize the number of lattice-sums
 in both  real space and k-space.
In the simulation, we use $\xi=\sqrt{\pi}/\bar{L}$
 where $\bar{L}$ is the average of the length of the periodic cell.
The suffix $\gamma=(n_1,n_2,n_3)$ represents
 a periodic cell in the real space
 and ${\bf r}_\gamma$ is the lattice vector given by
\begin{equation}
  {\bf r}_\gamma
  =
  \left(
  n_1L_1,
  n_2L_2,
  n_3L_3
  \right),
\end{equation}
where $L_1,L_2,L_3$ denotes the length of unit
 cell in each direction.
The wave vector ${\bf k}_\lambda$ in the reciprocal cell
 $\lambda=(m_1,m_2,m_3)$ is given by
\begin{eqnarray}
  &&{\bf k}_\lambda=
  \left(\hspace{.2cm}
  {2\pi m_1\over L_1},\hspace{.5cm}
  {2\pi m_2\over L_2},\hspace{.5cm}
  {2\pi m_3\over L_3}
  \hspace{.2cm}\right).
\end{eqnarray}
The summation $\mathop{{\sum}'}M^{(1)}_{ij}$
 in Eq.(\ref{eq:sim-ewald}) means that the contribution of
 the sum in which $\alpha = \beta$ at $\gamma = 0$ is eliminated.
The contribution of ${\bf k}=\vec{0}$ in the lattice sum
 in reciprocal space is canceled by that of the average force
 acting on the fluid\cite{brady1988b}.

We have checked the validity of our program to examine several sedimentation
 velocities under regular configurations like
 simple cubic, body-center cubic and face-center cubic.
We compared our results with the Stokesian dynamics by
 Brady {\it et al.}\cite{brady1988b} in which we choose the corresponding
 model, neglect higher order moment of hydrodynamic interactions,
 and with the exact solution by Zick and Homsy \cite{zick1982}.
 From this test,
 we recover the corresponding result by Brady {\it et al.}\cite{brady1988b},
 that is, our program seems to work correctly.
Furthermore,
 our result is very close to the exact solution by Zick and Homsy
 \cite{zick1982}
 except for extremely concentrated region.
Thus the simplification of neglecting higher order moments
 does not cause  serious problems.

\section{Fixed particles}
\label{sec:meth-fixed}

In this section,
 we show how to calculate the terminal velocities
 when we introduce particles fixed in the space.
If we obtain the terminal velocities,
 the motions of free particles are determined by (\ref{eq:time-evol-vel}).

If we know the resistance matrix
  which contains both free particles and fixed particles,
  we obtain the following equation,
\begin{equation}
  \left[
  \begin{array}{c}
    {\bf F}_m\\
    {\bf F}_f
  \end{array}\right]
  =
  \left[
  \begin{array}{cc}
    {\sf R}_{mm} &
    {\sf R}_{mf} \\
    {\sf R}_{fm} &
    {\sf R}_{ff}
  \end{array}\right]
  \cdot
  \left[
  \begin{array}{c}
    {\bf U}_m-{\bf u}^\infty_m \\
    {\bf U}_f-{\bf u}^\infty_f
  \end{array}\right]
  \label{eq:fix-res},
\end{equation}
where the suffix $m$ represents free particles and
  the suffix $f$ represents fixed particles.
We already know the force acting on free particles ${\bf F}_m$,
  the velocity of fixed particles ${\bf U}_f=\vec{0}$
  and the velocity of induced fluid ${\bf u}^\infty$.
(\ref{eq:fix-res}) can be solved for ${\bf U}_m$ as
\begin{equation}
  {\bf U}_m-{\bf u}^\infty_m
  =
  {\sf R}_{mm}^{-1}
  \cdot
  \left(
  {\bf F}_m
  +
  {\sf R}_{mf}
  \cdot
  {\bf u}^\infty_f\right).
\end{equation}
Therefore the terminal velocities of free particles can ben
  represented by known variables.
This procedure is applicable to systems with periodic boundary conditions.
In fact, from Eq.(\ref{eq:sim-ewald}) the mobility matrix can be composed
 of particles in the unit cell.



\begin{figure}
  \caption{
    The relationship between particle radius and dimensionless
    parameters.
    The straight lines connects the points in which dimensionless
    parameters are 1.
    }
  \label{fig:param}
\end{figure}

\begin{figure}
  \caption{
    Snapshots of the slug.
    The number of particles in the unit cell is 72 and
    the volume fraction is $0.348$.
    The ratio of the height to the width is 6.
    Time interval is $2.0\times 10^{-2}$ sec.
    }
  \label{fig:slug}
\end{figure}

\begin{figure}
  \caption{
    Snapshots of the slug in three-dimensional simulation.
    The number of particles is 50 in the unit cell and
    the volume fraction is $0.45$.
    The ratio of the height to the width is 4,
    and time interval between them is $2.0\times 10^{-2}$ [sec].
    }
  \label{fig:slug-3d}
\end{figure}

\begin{figure}
  \caption{
    A snapshot of the simulation with square cells
    with the periodic boundary condition.
    It is shown with four periodic images.
    The number of particles in the unit cell is 90 and
    the volume fraction is $0.327$.
    }
  \label{fig:square}
\end{figure}

\begin{figure}
  \caption{
    The sequence of the simulation with fixed particles.
    The number of free particles and fixed particles
    in the unit cell are 128 and 10, respectively.
    The velocity of induced fluid is $u^\infty = 0.3 U_0$,
    where $U_0 = m\tilde{g}/6\pi\mu a$ is
    the one particle sedimentation velocity.
    The time interval from the left-up to the right-down is $0.112$ sec.
    We can see three bubbles flowing up through the bed
    (except for the initial instabilities).
    }
  \label{fig:fixed}
\end{figure}

\begin{figure}
  \caption{
    The time evolution of standard deviation of particle velocities
    in the simulation of Fig. \protect\ref{fig:fixed}.
    Velocities are normalized by one particle sedimentation velocity $U_0$.
    1 step means $10^{-4}$ sec.
    }
  \label{fig:fixed-variance}
\end{figure}

\begin{figure}
  \caption{
    The time evolution of the volume fraction of particles
    in the area that
    is the horizontal region at 8/25 of the cell height above the bottom
    in the simulation of Fig. \protect\ref{fig:fixed}.
    1 step means $10^{-4}$ seconds.
    }
  \label{fig:fixed-density}
\end{figure}

\begin{figure}
  \caption{
    The velocity distribution functions (VDFs) of
    (a) the horizontal direction
    and (b) the vertical direction
    in the simulation of Fig. \protect\ref{fig:square}
    (with square unit cells).
    The velocities are normalized by
    one particle sedimentation velocity $U_0$.
    The vertical component of velocity is plotted
    in the coordinate system in which the average velocity of the fluid is zero
    and the direction of gravity is negative direction.
    The horizontal VDF can be approximated by
    $\exp (-20.0\hat{U}^2)$, where $\hat{U}\equiv U/U_0$.
    The left hand side and the right hand side of the vertical VDF
    can be approximated by
    $\exp (-10.0(\hat{U}+0.2)^2)$
    and $\exp (-4.5(\hat{U}+0.2)^2)$, respectively.
    The VDF $f(\hat{U})$ is normalized by $\int d\hat{U}\ f(\hat{U}) = 1$.
    }
  \label{fig:veldist-square}
\end{figure}

\begin{figure}
  \caption{
    The velocity distribution functions (VDFs)
    in the simulation of
    Fig. \protect\ref{fig:fixed}(with fixed particles).
    The velocities are normalized by
    one particle sedimentation velocity $U_0$
    and we use the coordinate system
    in which the fixed particle's velocity become zero.
    The VDFs in horizontal direction and vertical direction
    can be approximated by
    $\exp (-5.0|\hat{U}|)$ and $\exp (-3.6|\hat{U}|)$, respectively.
    Here $\hat{U}\equiv U/U_0$.
    The normalization is the same as that
    in Fig. \protect\ref{fig:veldist-square}.
    }
  \label{fig:veldist-fixed}
\end{figure}

\begin{figure}
  \caption{
    The velocity distribution function (VDF) in (a) the active state
    and (b) the inactive state of the simulation with fixed particles
    which is the same of that in Fig. \protect\ref{fig:veldist-fixed}.
    The velocities are normalized by
    one particle sedimentation velocity $U_0$
    and we use the coordinate system
    in which the fixed particle's velocity become zero.
    The Gaussian-fitted function is also ploted.
    }
  \label{fig:fixed-vdf-hard-soft}
\end{figure}

\begin{figure}
  \caption{
    The velocity distribution function (VDF) of
    (a) the horizontal direction
    and (b) the vertical direction
    in the simulation of Fig. \protect\ref{fig:slug}
    (with rectangular unit cells).
    The velocities are normalized by
    one particle sedimentation velocity $U_0$.
    The VDF in the horizontal direction can be approximated by
    $\exp (-25.0|\hat{U}|)$ in the exponential form or
    $(1.0+700.0\hat{U}^2)^{-1.5}$ in the t-distribution form.
    The VDF in the vertical direction for the right-hand-side by
    can be approximated by
    $\exp (-50.0|\hat{U}+0.115|)$, in the exponential form
    or by
    $(1.0+800.0(\hat{U}+0.115)^2)^{-2.0}$ in the t-distribution form.
    Here $\hat{U}\equiv U/U_0$.
    The normalization is the same as that
    in Fig. \protect\ref{fig:veldist-square}.
    }
  \label{fig:veldist-slug}
\end{figure}

\begin{figure}
  \caption{
    The substracted velocity distribution function (VDF)
    from the VDF in Fig. \protect\ref{fig:veldist-slug} (b).
    We make this figure by substracting the uniform distribution,
    representing the contribution of falling particles in the slug,
    from the original VDF in the range of $-0.4<\hat{U}<-0.115$.
    This VDF can be fitted
    for the left-hand-side by
    $(1.0+700.0(\hat{U}+0.115)^2)^{-1.5}$
    and for the right-hand-side by
    $(1.0+800.0(\hat{U}+0.115)^2)^{-2.0}$.
    Here $\hat{U}\equiv U/U_0$.
    }
  \label{fig:vdf-slug-sub}
\end{figure}

\begin{figure}
  \caption{
    Power spectrum $E(\omega)$ of
    (a) the holizontal velocity and (b) the vertical velocity
    in the simulation of Fig. \protect\ref{fig:square}
    (with square unit cells).
    The frequency is normalized by
    $\omega_0=2\pi/2048\ {\rm[step^{-1}]} \simeq 30.7 {\rm [sec^{-1}]}$.
    The straight lines are least-square fits between the frequency from
    $\omega_{r}$ to $\omega_{p}$ with $-1.597\pm 0.009$
    and $-1.626\pm 0.009$ slope for $U_x$ and $U_y$ respectively.
    }
  \label{fig:pow-t-square}
\end{figure}

\begin{figure}
  \caption{
    Power spectrum $E(\omega)$ of
    (a) the holizontal velocity and (b) the vertical velocity
    in the simulation of Fig. \protect\ref{fig:slug}
    (with rectangular unit cells).
    The frequency is normalized by
    $\omega_0=2\pi/2048\ {\rm[step^{-1}]} \simeq 30.7 {\rm [sec^{-1}]}$.
    The straight lines are least-square fits between the frequency from
    $\omega_{r}$ to $\omega_{p}$ with $-1.494\pm 0.01$
    and $-1.513\pm 0.009$ slope for $U_x$ and $U_y$ respectively.
    }
  \label{fig:pow-t-slug}
\end{figure}

\begin{figure}
  \caption{
    Power spectrum $E(\omega)$ of
    (a) the holizontal velocity and (b) the vertical velocity
    in the simulation of Fig. \protect\ref{fig:fixed}(with fixed particles).
    The frequency is normalized by
    $\omega_0=2\pi/2048\ {\rm[step^{-1}]} \simeq 30.7 {\rm [sec^{-1}]}$.
    The straight lines are least-square fits between the frequency from
    $\omega_{r}$ to $\omega_{p}$ with $-1.514\pm 0.007$
    and $-1.490\pm 0.007$ slope for $U_x$ and $U_y$ respectively.
    }
  \label{fig:pow-t-fixed}
\end{figure}


\begin{table}
  \caption{
    Values of the dimensionless parameters and
    the characteristic time $T_{r}$[sec] and $T_{p}$[sec]
    when the particle radius is $10^{-3}$[cm].
    }
  \label{tab:param}
  \begin{tabular}{|c|c|c|}
    &Air & Water\\
    \hline
    $St$&$9.1 \times 10^0$ & $1.8 \times 10^{-3}$   \\
    $Pe$&$2.5 \times 10^5$ & $1.5 \times 10^5$   \\
    $Re$&$2.0 \times 10^{-2}$ & $3.3 \times 10^{-3}$   \\
    \hline
    $T_{r}$&$3.1 \times 10^{-3}$ & $5.6 \times 10^{-5}$   \\
    $T_{p}  $&$3.3 \times 10^{-4}$ & $3.1 \times 10^{-2}$   \\
  \end{tabular}
\end{table}

\begin{table}
  \caption{
    Values of $Re_{(p)}$ in our simulations,
    where $L$ is measured by the maximum length of the unit cell.
    }
  \label{tab:froude}
  \begin{tabular}{|c|c|c|c|}
    &square cell(Fig. \ref{fig:square})
    &rectangular cell(Fig. \ref{fig:slug})
    &with fixed particles(Fig. \ref{fig:fixed})\\
    \hline
    $Re_{(p)}$& $9.9\times 10^2$ & $2.1\times 10^3$ & $2.1\times 10^3$ \\
    $Re_{(p)}^{3/4}$& $1.8\times 10^{2}$ & $3.1\times 10^{2}$
    & $3.1\times 10^{2}$ \\
  \end{tabular}
\end{table}

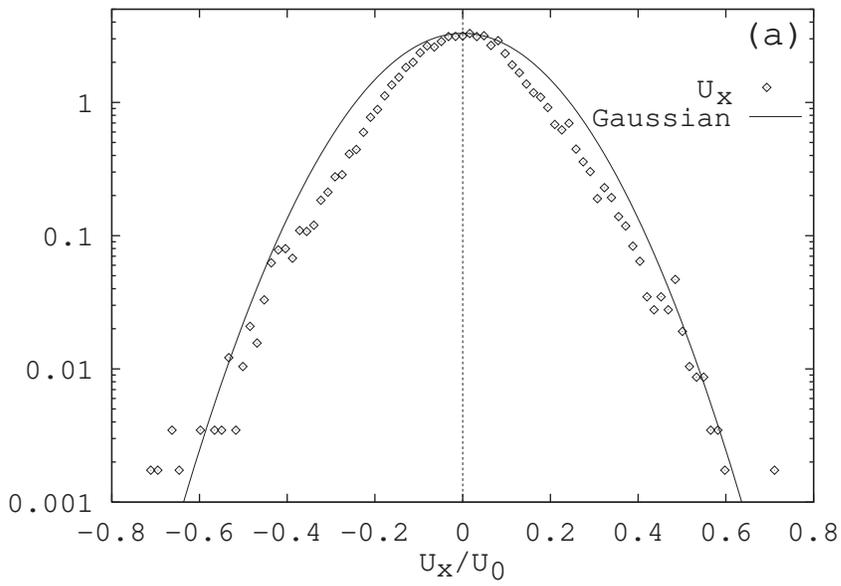
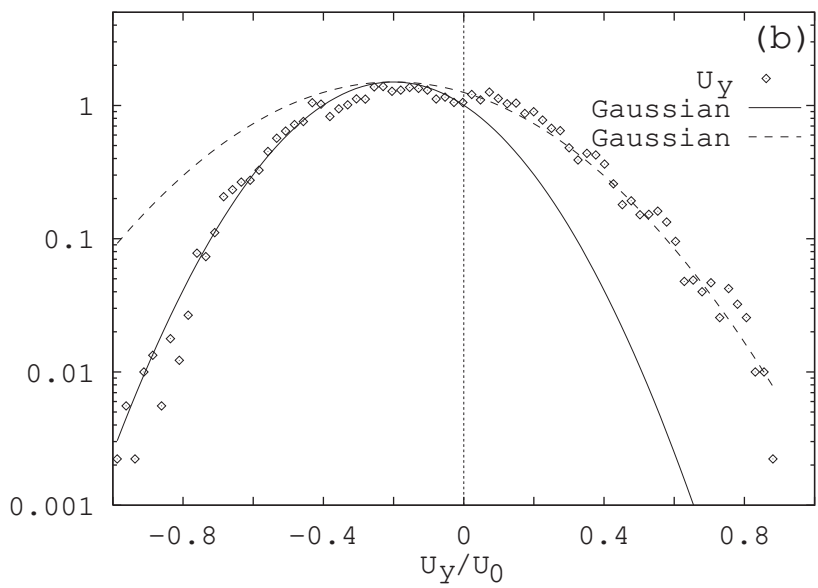
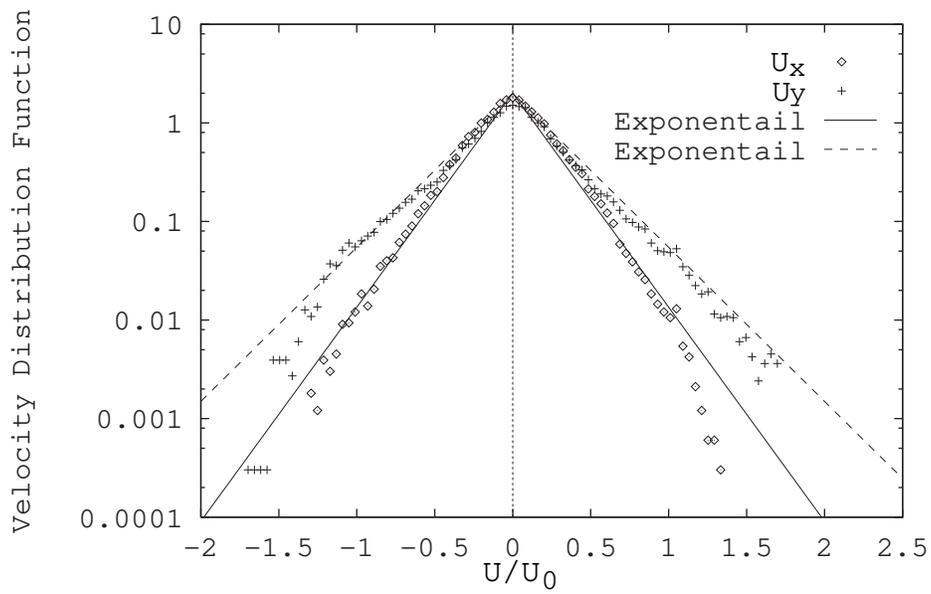
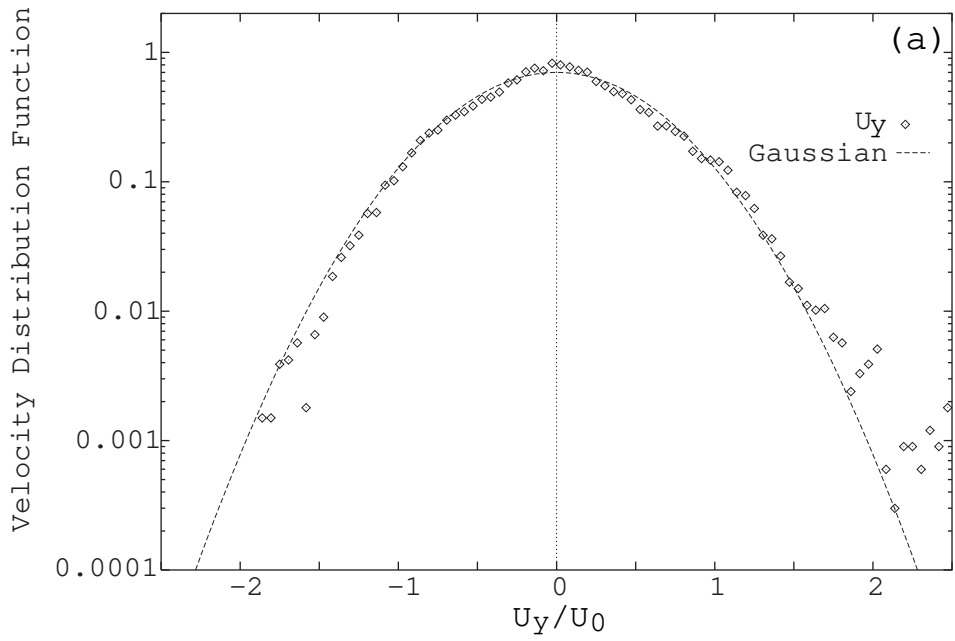
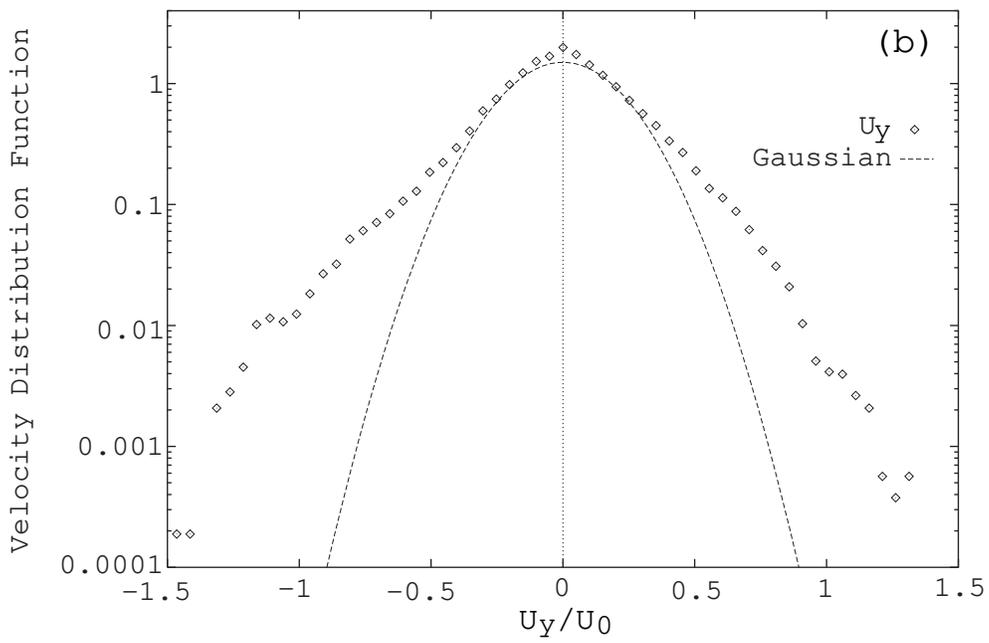
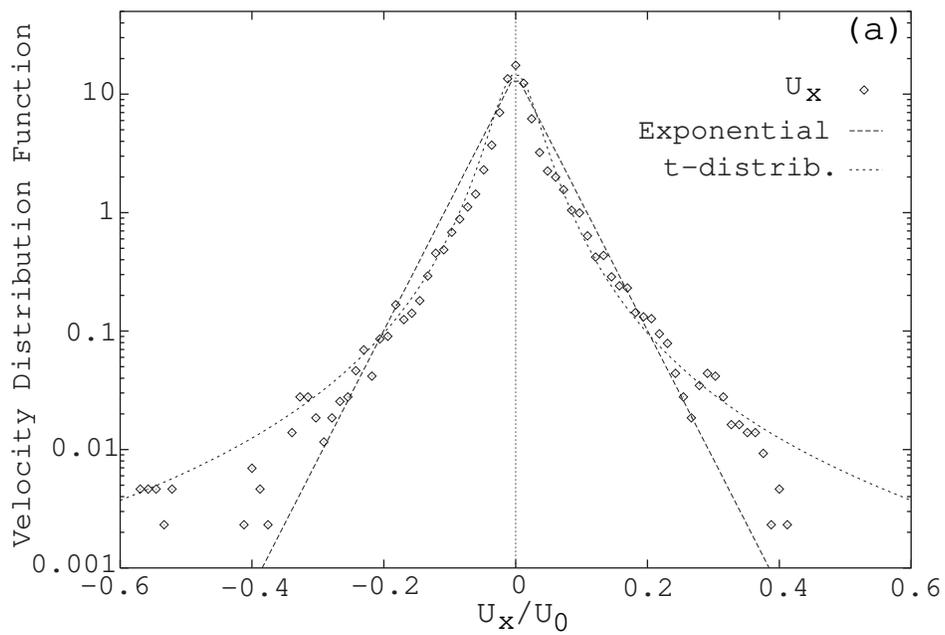
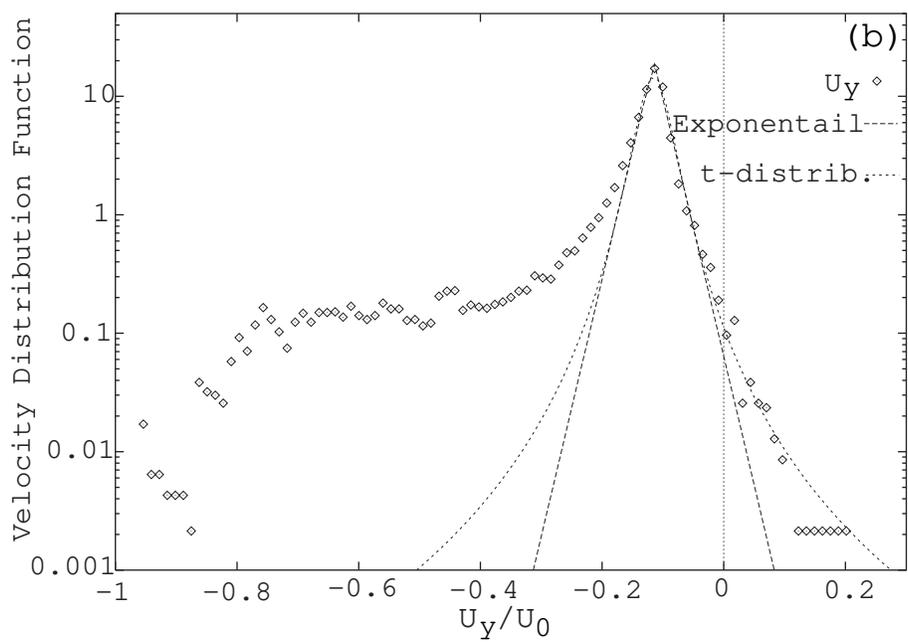
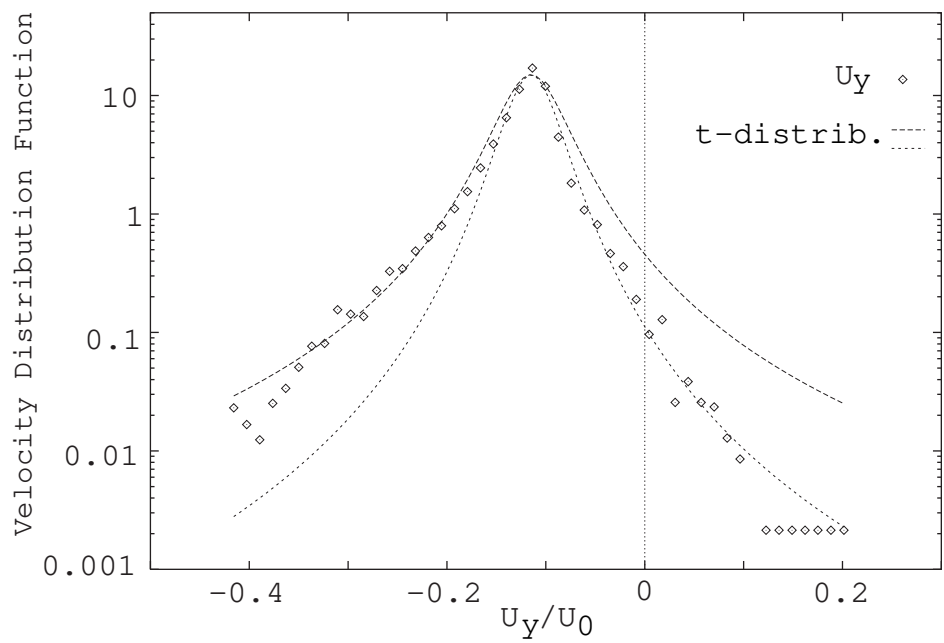
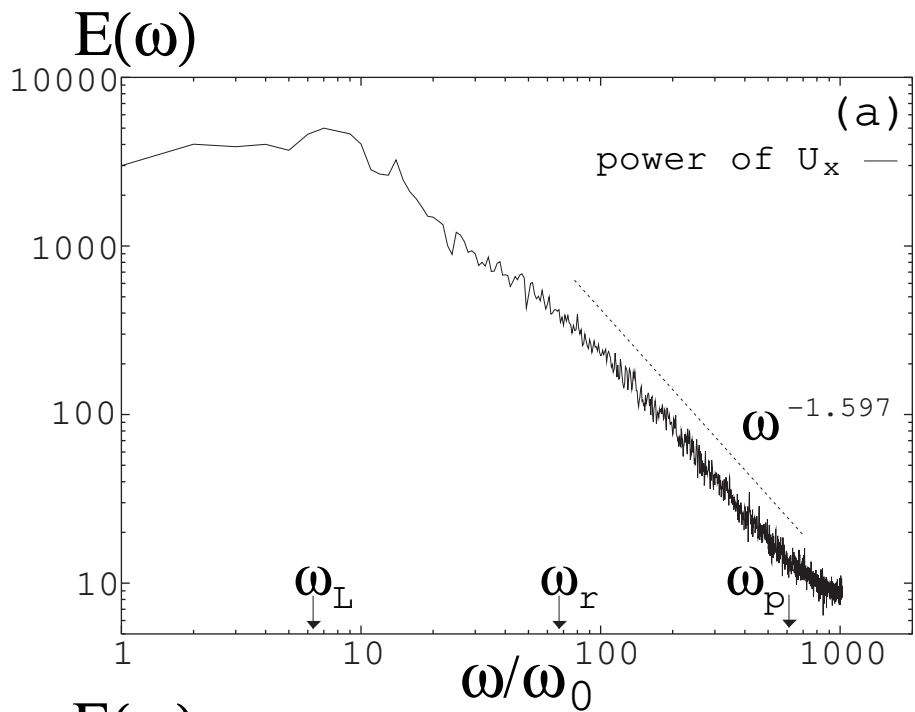
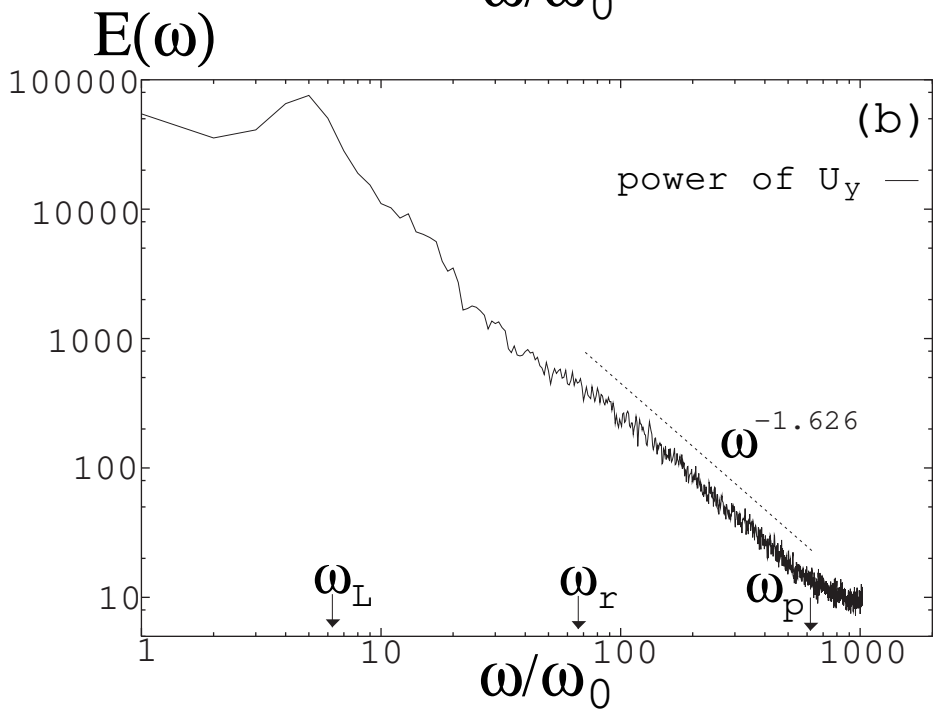
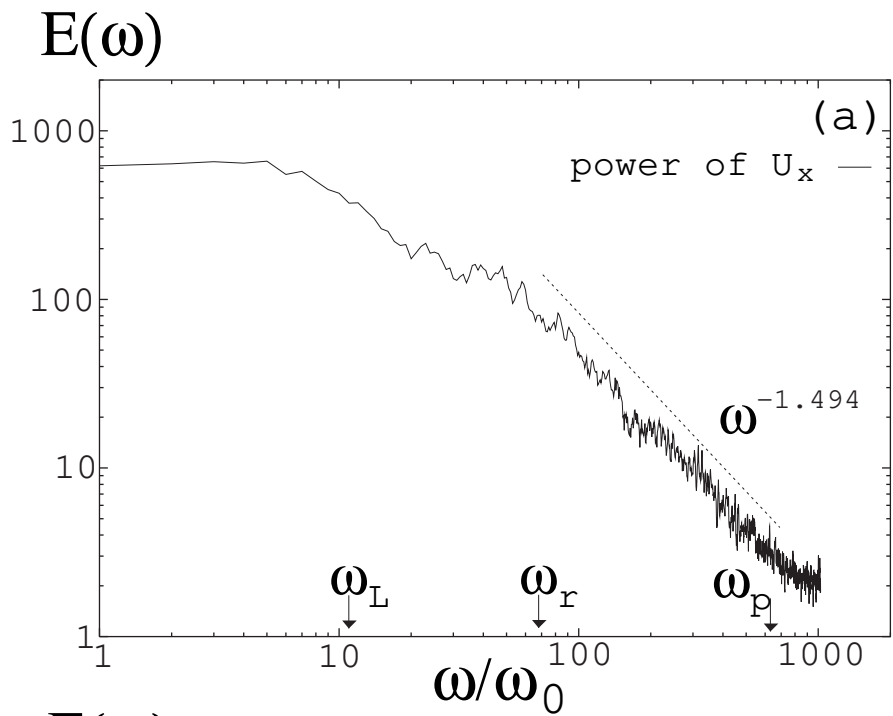
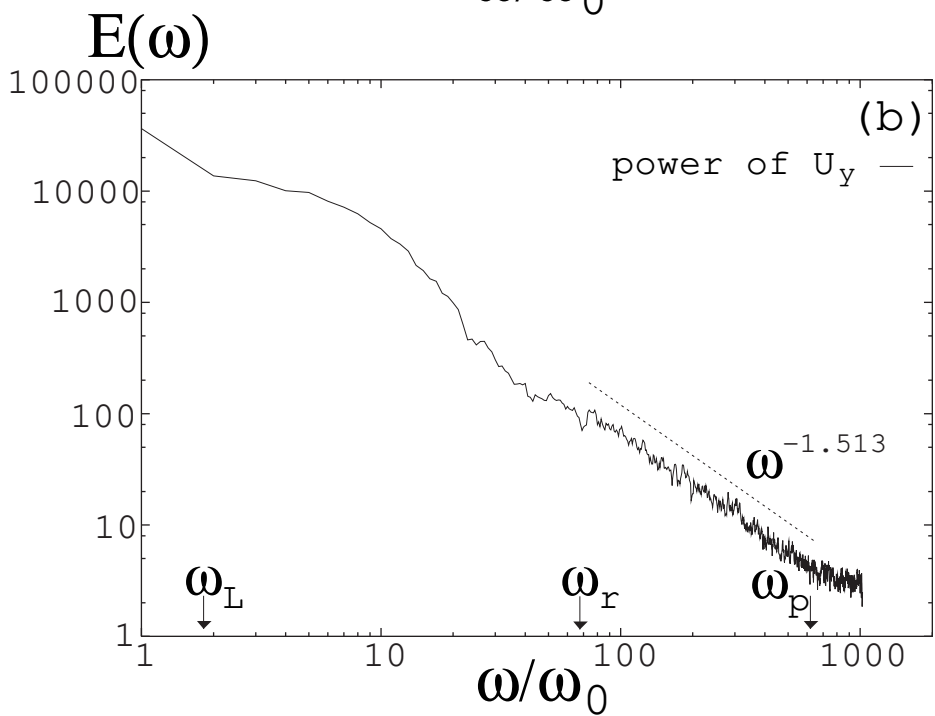
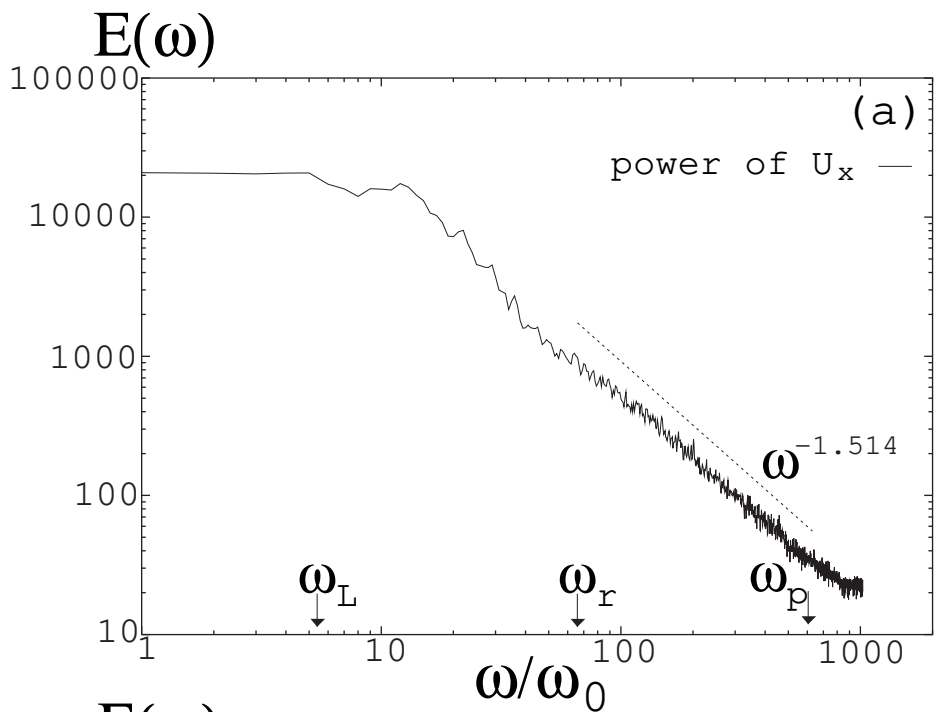
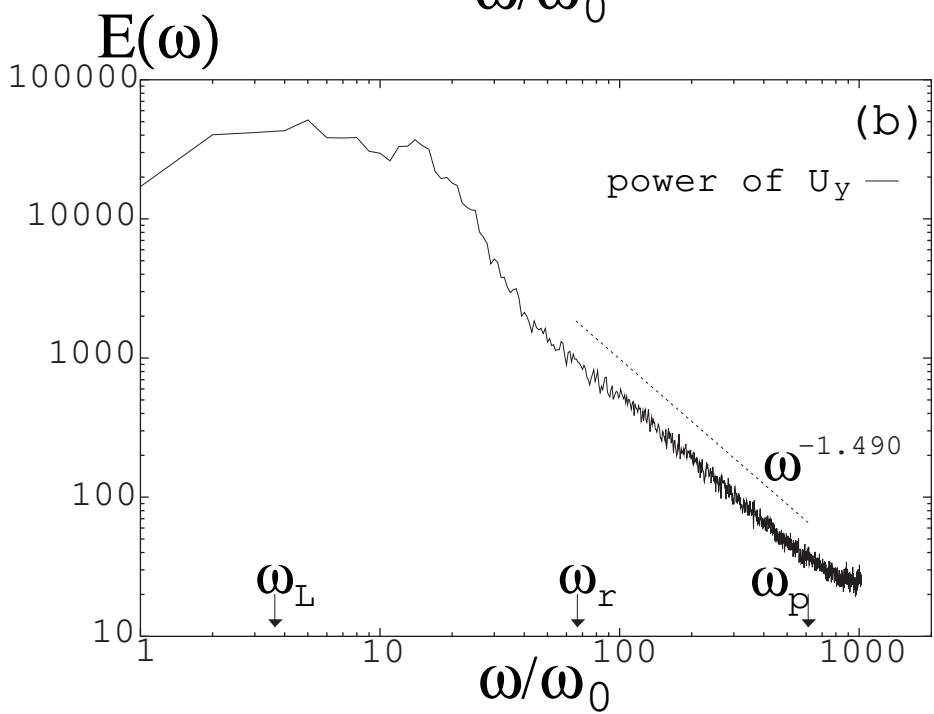
\end{document}